\documentclass[prb, 11pt, aps, showpacs, longbibliography,preprint, nofootinbib, endfloats*]{revtex4-2}

\usepackage[dvipsnames]{xcolor}
\usepackage{amssymb,bm}
\usepackage{graphicx}
\usepackage{amsmath} 
\usepackage{setspace}
\usepackage{enumerate}
\usepackage{bbold}
\usepackage{esint}
\usepackage{float}
\usepackage{euscript}
\usepackage[bookmarks, colorlinks=true, breaklinks]{hyperref}
\hypersetup{linkcolor=blue,citecolor=blue,filecolor=black,urlcolor=blue}
\usepackage[normalem]{ulem}

\DeclareMathOperator{\Tr}{Tr}
\newcommand{\norm}[1]{\left| {#1} \right|}

\begin{document}

\title{Landau Theory of Barocaloric Plastic Crystals}

\author{R. Marín-Delgado}
\affiliation{$^{1}$Centro de Investigaci\'{o}n en Ciencia e Ingenier\'{i}a de Materiales, Universidad de Costa Rica, Costa Rica}
\affiliation{$^{2}$Escuela de F\'{i}sica, Universidad de Costa Rica, Costa Rica}

\author{X. Moya\footnote{xm212@cam.ac.uk}}
\affiliation{$^{3}$Department of Materials Science, University of Cambridge, United Kingdom}

\author{G. G. Guzm\'{a}n-Verri\footnote{gian.guzman@ucr.ac.cr}}
\affiliation{$^{1}$Centro de Investigaci\'{o}n en Ciencia e Ingenier\'{i}a de Materiales, Universidad de Costa Rica, Costa Rica}
\affiliation{$^{2}$Escuela de F\'{i}sica, Universidad de Costa Rica, Costa Rica}
\affiliation{$^{3}$Department of Materials Science, University of Cambridge, United Kingdom}

\date{\today}


\begin{abstract}
We present a minimal Landau theory of plastic-to-crystal phase transitions in which the key components are  a multipole-moment order parameter that describes the orientational ordering of the constituent molecules,  coupling between such order parameter and elastic strains, and thermal expansion. We illustrate the theory with the simplest non-trivial model in which the orientational ordering is described by a quadrupole moment, and use such model to calculate barocaloric effects in plastic crystals that are driven by hydrostatic pressure. The model captures characteristic features of plastic-to-crystal phase transitions, namely, large changes in volume and entropy  at the transition, as well as the linear dependence of the transition temperature with pressure. We identify temperature regions in the barocaloric response associated with the individual plastic and crystal phases, and those involving the phase transition. Our model is in overall agreement with previous experiments in powdered samples of fullerite C$_{60}$, and 
predicts peak isothermal entropy changes of $\sim90 \,{\rm J K^{-1} kg^{-1}}$ and peak adiabatic temperature changes of $\sim35 \,{\rm K}$ under $0.60\,$GPa at $265\,$K in fullerite single crystals.
\end{abstract}
\maketitle

\section{Introduction}

Barocaloric (BC) effects are reversible thermal changes in response to changes in applied hydrostatic pressure and are usually parametrized by isothermal changes in entropy or adiabatic changes in temperature~\cite{Moya2020a}. The current need for cooling and heating technologies that are free of greenhouse gases has led to a resurgence of BC studies, which date back to the late 20th century~\cite{Moya2014a}. While large BC effects~($ \lesssim 100\,{\rm J K^{-1} kg^{-1}}$) have been observed in various materials such as oxides, metals, and superionic conductors~\cite{Moya2020a, Moya2014a, Lloveras2021a, Boldrin2021a, Cirillo2022a}, only recently have colossal BC effects~($\gtrsim 100\,{\rm J K^{-1} kg^{-1}}$) been identified in a class of orientationally disordered solids known as plastic crystals (PCs)~\cite{Moya2018a}. These colossal BC effects are comparable to the pressure-driven thermal changes seen in hydrofluorocarbons/hydrocarbons used in vapor-compression refrigeration. PCs exhibiting colossal BC effects include neopentane derivatives~\cite{ Lloveras2019a, Li2019a, Aznar2020a, Salvatori2023b}, adamantane~\cite{Meijer2023a}, adamantane derivatives~\cite{Aznar2021a, Salvatori2022a}, carboranes~\cite{Zhang2022a}, and alkali hydrido-closoborates~\cite{Zeng2024a}. In addition, orientationally disordered solids which are not as mechanically soft as PCs, can also display large~\cite{Junning2020a, Ren2022a} and colossal~\cite{Zhang2023a} BC effects. Such materials include fullerite~\cite{Junning2020a}, ammonium salts~\cite{Ren2022a, Zhang2023b, Gao2024a}, and hexafluophosphates~\cite{Zhang2023a}, and are typically discussed alongside PCs~\cite{Sherwood1979a, Lloveras2023b}.

PCs are solids made of molecules with an orientational degree of freedom and whose centres of mass form a lattice with long-range positional order~\cite{Das2020a}. In the parent high-temperature plastic phase, some or all of the constituent molecules perform nearly uncorrelated rotations, forming ``molecular globules", and rendering the solid orientationally disordered~\cite{Sherwood1979a}. In the low-temperature crystal phase, the molecules spontaneously align along one or several axes, thus breaking rotational symmetry. Such changes in the orientational order are usually accompanied by very large latent heats and change in volume, which lead to very large BC effects when driving the transition with pressure. 

The discovery of colossal BC effects in PCs has sparked theoretical and computational interest primarily using density functional theory~\cite{Li2020a}, molecular dynamics methods~\cite{Li2022a,  Zeng2024a, Li2023a, Escorihuela-Sayalero2024a}, supercell lattice dynamics calculations~\cite{Meijer2023a}, and mean-field microscopic modelling~\cite{deOliveira2023a}, aimed at understanding  plastic-to-crystal transitions and predicting their associated BC response. However, a macroscopic description based on Landau theory is missing. The purpose of this work is to propose such a theory, and illustrate it with a minimal model that captures the essential features of plastic-to-crystal phase transitions. The main advantage of this approach is that both the molecular and lattice symmetries are incorporated into the model by construction, as well as the couplings between the order parameter (OP) of the transition and the elastic strains. We consider the simplest non-trivial case in which the orientational ordering is described by a quadrupolar OP, as is often done for nematic ordering in liquid crystals. Consequently, our model draws parallels to the Landau-de Gennes theory of phase transitions in liquid crystals~\cite{Selinger2015a}.

The paper is organized as follows: first, we introduce  the orientational OP of plastic-to-crystal phase transitions, we develop the relevant Landau free energy density expansion, perform its minimization, and derive expressions for the BC properties (Methods section).  Second, we fit our model to experimental data in solid C$_{60}$ and provide a comparison with measured BC quantities (Results and Discussion). Finally, we summarize our findings and discuss the limitations of our model and its possible extensions (Summary and Conclusions). 

\section{Methods}

\subsection{Orientational and strain order parameters}

Spontaneous long-range orientational orderings such as those occurring at  plastic-to-crystal phase transitions can be described by multipole-moment OPs consistent with both the rotation and inversion symmetries of the constituent molecules~\cite{Lynden-Bell1994a, Meyer1978a}. We consider the simplest possible non-trivial case in which the molecules do not exhibit inversion symmetry, i.e. a quadrupole moment ${\bm Q}$. In addition, we consider the elastic strain ${\bm \epsilon}$ that results from the orientational ordering,  hydrostatic pressure, and the thermal expansion.

\subsection{Free-energy density} \label{sec:free-energy}

We consider an isotropic solid  subjected to a hydrostatic pressure $P$, and near a plastic-to-crystal transition with a homogeneous quadrupole ${\bm Q}$, and a homogeneous strain ${\bm \epsilon}$. We propose a free-energy density $G$ given as follows,
\begin{align}
    \label{eq:freeenergy}
    G=G_0+G_Q+G_\epsilon + G_{\epsilon Q}+P \Tr {\bm \epsilon},
\end{align}
where $G_0=G_0(T,P)$ is the free-energy density of the background, $G_Q$ is the rotational energy,
\begin{align*}
    G_Q=a\left(T-T_Q\right)\Tr {\bm Q}^2+b\Tr{\bm Q}^3+c\left(\Tr{\bm Q}^2\right)^2,
\end{align*}
$G_\epsilon$ is the strain energy, which includes thermal expansion,
\begin{align*}
G_\epsilon=-K\alpha\left(T-T_0\right)\Tr{\bm \epsilon}+\frac{1}{2}\left( K - 2\mu /3 \right)\left(\Tr{\bm \epsilon}\right)^2+\mu\Tr{\bm \epsilon}^2,
\end{align*}
and $G_{\epsilon Q}$ is the strain-orientation coupling energy, 
\begin{align*}
     G_{\epsilon Q}= d\Tr\left({\bm \epsilon} \cdot {\bm Q}\right) +(g-e/3) \left( \Tr{\bm \epsilon} \right) \left(\Tr{\bm Q}^2\right)  +e\Tr\left( {\bm \epsilon} \cdot {\bm Q}^2\right). 
\end{align*}
The parameters $a, b, c, d, e$ and $g$ are independent of temperature and pressure, and $a$ and $c$ are assumed to be positive; $T_Q$ is the limit of stability of the symmetric phase in the absence of strain-orientation coupling; and $\alpha, K$ and $\mu$ are, respectively, the volumetric coefficient of thermal expansion in the parent phase, the bare bulk modulus, and the bare shear modulus.
We define the undeformed state as the volume of the material at atmospheric pressure at temperature $T_0$~\cite{LandauElasticity}.
We then set the value of this reference undeformed volume equal to one, and normalize relevant thermodynamic quantities with respect to this  volume. We note that the elastic strains ${\bm \epsilon}$ are generated by the acoustic modes of the solid~\cite{Ashcroft1976a}, which implies that such vibrations are implicitly incorporated into our model. 

At equilibrium, the strain ${\bm \epsilon}$ is given by the minimization $\partial G / \partial {\bm \epsilon}=0$~\cite{Corrales2017a, Cai2020a}, which gives the following result,
\begin{align}
    \label{eq:strain}
    {\bm \epsilon}=\frac{\mathbb{1}}{3}\left[\alpha \left(T-T_0\right)-\frac{P}{K}+\left(\frac{e}{2\mu}-\frac{g}{K}\right)\Tr\bm{Q}^2\right] - \frac{d}{2\mu}\bm{Q} - \frac{e}{2\mu}\bm{Q}^2.
\end{align}
We then substitute Eq.~(\ref{eq:strain}) into Eq.~(\ref{eq:freeenergy}), which yields,
\begin{multline}
    \label{eq:Gtilde}
    G=G_0+\left[\tilde{a}\left(T-\tilde{T}_Q\right)  -\frac{g}{K}P \right]\Tr{\bm Q}^2+\tilde{b}\Tr{\bm Q}^3+\tilde{c}\left(\Tr{\bm Q}^2\right)^2 \\ -\frac{1}{2K} \left[P - \alpha K \left(T-T_0\right)\right]^2,
\end{multline}
where $\tilde{a}, \tilde{b}$, $\tilde{c}$ and $\tilde{T}_Q$ are renormalized parameters given as follows,
\begin{align*}
    \tilde{a}&=a+g\alpha,\\
    \tilde{b}&=b-\frac{de}{2\mu},\\
    \tilde{c}&=c - \frac{1}{2}\left(\frac{g^2}{K}+\frac{e^2}{12\mu}\right), \\
     \tilde{T}_Q  &= \left(T_Q+\frac{d^2}{4\mu a} + \frac{g\alpha}{a} T_0\right) \left(1+\frac{g \alpha}{a}\right)^{-1}.
\end{align*}
We have thus arrived at a free-energy density that, in the absence of pressure and thermal expansion, shares an identical form to the Landau-de Gennes theory of isotropic-to-nematic phase transitions in liquid crystals~\cite{Selinger2015a}, and therefore, exhibits similar features, the most important of which is the first-order character of the phase transition due to the cubic invariant $\tilde{b}\Tr{\bm Q}^3$. 

Equation~(\ref{eq:Gtilde}) is the starting point for finding the equilibrium configuration ${\bm Q}(T,P)$, and from which we compute the relative volume,
\begin{align}
    \label{eq:Tre}
   V(T,P) = \left(\frac{\partial G}{\partial P}\right)_T = V_0 +   \alpha(T-T_0) - \frac{P}{K} - \frac{g}{K} \Tr{\bm Q}^2(T,P),
\end{align}
and the entropy density,
\begin{align}
    \label{eq:entropy}
   S(T,P)= - \left(\frac{\partial G}{\partial T}\right)_P =S_0  +\alpha^2 K(T-T_0) -\alpha P -\tilde{a} \Tr{\bm Q}^2(T,P),
\end{align}
where $V_0= \left(\partial G_0/ \partial P\right)_T$ 
and $S_0 = - \left(\partial G_0/ \partial T\right)_P$ are the relative volume and entropy density associated with the background, respectively. The background $G_0$ is an input to the model, which we will discuss in Sec.~\ref{sec:background}.

\subsection{Plastic-to-crystal phase transition} \label{sec:plastic-to-crystal}

We consider a plastic-to-crystal phase transition in which the orientational ordering occurs along a single axis $\hat{\bm n}=(0,0,1)$. The quadrupole is thus given as follows~\cite{Selinger2015a},
\begin{align}
    \label{eq:C60OP}
   {\bm Q} =  \frac{3}{2} A \left( \hat{\bm n} \otimes \hat{\bm n} - \frac{1}{3}  \mathbb{1} \right)= \begin{pmatrix} -A/2 & 0 & 0 \\ 0 &   -A/2  & 0 \\ 0 & 0 & A \end{pmatrix},
\end{align}
where $A$ is a variational parameter that determines the degree of molecular alignment around $\hat{\bm n}$.  In the plastic phase, the orientation of the molecules is random, thus $A=0$. In the crystal phase, the molecular rotations are correlated, and tend to align along $\hat{\bm n}$, which leads to $A > 0$. The case $A<0$ involves an ordered phase in which the molecules prefer to align perpendicular to $\hat{\bm n}$, but such phase turns out to be metastable at all temperatures and pressures.  

Substitution of Eq.~(\ref{eq:C60OP}) into  Eq.~(\ref{eq:Gtilde}) gives the following free-energy density,
\begin{align}
    \label{eq:renormalizedG}
G=G_0+\frac{1}{2}\left[\tilde{a}\left(T-\tilde{T}_Q\right)  -\frac{g}{K}P \right]A^2-\frac{\tilde{b}}{3} A^3+\frac{\tilde{c}}{4}A^4-\frac{1}{2K} \left[P - \alpha K (T-T_0)\right]^2,
\end{align}
where we have rescaled the model parameters of Eq.~(\ref{eq:Gtilde}) as follows: $\tilde{a} \to \tilde{a}/3 $, $g \to g/3$, $\tilde{b} \to - 4 \tilde{b}/9 $, and $\tilde{c} \to  \tilde{c}/9 $.  By minimizing Eq.~(\ref{eq:renormalizedG}) with respect to $A$, we obtain the following result,
\begin{align}
    \label{eq:Aexpansion}
    A\left(T,P\right) = \begin{cases} 0,& T \geq T_c, \\ \left(2\tilde{b}/3\tilde{c}\right) \left[ 1 +\left(9 \tilde{a} \tilde{c}/2 \tilde{b}^2\right)\left(T_c-T\right) \right], & T \leq T_c, \end{cases}
\end{align}
where $T_c$ is the pressure-induced transition temperature,
\begin{align}
    \label{eq:Tc}
    T_c&=T_c^0+ \left(\frac{1}{\tilde{a}}\frac{g}{K}\right)P,
\end{align}
and $T_c^0 = \tilde{T}_Q+\frac{2\tilde{c}\tilde{b}^2}{9\tilde{a}} $ is the transition temperature at ambient pressure.  We note that in deriving Eq.~(\ref{eq:Aexpansion}), we have expanded $A(T,P)$ to linear order in $(T_c-T)$ for $ T \leq T_c$. 

By setting $T_0=T_c^0$ and substituting Eqs.~(\ref{eq:C60OP}) and~(\ref{eq:Aexpansion}) into Eq.~(\ref{eq:Tre}), we obtain the relative volume,
\begin{align}
    \label{eq:volume}
 V(T,P) =
 \begin{cases}
  V_0+ \alpha(T-T_c) - \left( 1-  K \alpha \, dT_c/dP   \right)  \frac{P}{K}, & T \geq T_c, \\
   V_0+ \left(\alpha + \Delta \alpha \right)  (T-T_c)  - \left(1-  K \alpha \, dT_c/dP \right) \frac{P}{K} + \Delta V_t, & T \leq T_c,
 \end{cases}
\end{align}
where $\alpha + \Delta \alpha$ is  the coefficient of thermal expansion in the ordered phase with $\Delta \alpha$ given as follows,
\begin{align}
    \label{eq:Dalpha}
    \Delta \alpha = \frac{2\tilde{a}}{\tilde{c}} \frac{g}{K},
\end{align}
$\Delta V_t$ is the change in relative volume at the phase transition,
\begin{align}
    \label{eq:DVt}
   \Delta V_t= - \frac{g}{2K} \Delta A_t^2,
\end{align}
with $\Delta A_t$ being the change in $A(T,P)$ at $T_c$,
\begin{align}
    \label{eq:OPjump}
    \Delta A_t = \frac{2}{3}\frac{\tilde{b}}{\tilde{c}},
\end{align}
and,
\begin{align}
    \label{eq:dTcdP}
  \frac{dT_c}{dP} =\frac{1}{\tilde{a}}\frac{g}{K}.
\end{align}

Substitution of Eqs.~(\ref{eq:C60OP}) and~(\ref{eq:Aexpansion}) into Eq.~(\ref{eq:entropy}), gives the entropy density,
\begin{align}
    \label{eq:entropy2}
   S(T,P) = \begin{cases}  
  S_0 + K \alpha^2 \left( T-T_c \right) - \left( 1- K \alpha \, dT_c/dP \right) \alpha P, & T \geq T_c, \\
   S_0 + \left(K \alpha^2+ \frac{\Delta \alpha}{dT_c/dP} \right) \left( T-T_c \right)- \left( 1- K \alpha \, dT_c/dP \right) \alpha P + \Delta S_t, & T \leq T_c,
     \end{cases}
\end{align}
where $\Delta S_t$ is the change in entropy at the phase transition,
\begin{align}
    \label{eq:DSt}
    \Delta S_t 
    =- \frac{\tilde{a}}{2} \Delta A_t^2.
\end{align}

We note $\Delta V_t$, $\Delta S_t$ and $dT_c/d P$ are not independent as they obey the Clausius-Clapeyron equation,
\begin{align}
    \label{eq:CC}
    \frac{dT_c}{d P} = \frac{\Delta V_t}{\Delta S_t},
\end{align}
which we recover from Eqs.~(\ref{eq:DVt}), (\ref{eq:dTcdP}) and (\ref{eq:DSt}). 

From Eq.~(\ref{eq:entropy2}), we obtain the volumetric heat capacity,
\begin{align}
    \label{eq:specific heat capacity}
   C(T,P) = T \left( \frac{\partial S}{\partial T}  \right)_P = \begin{cases}  
  C_0 + K \alpha^2 T, & T \geq T_c, \\
   C_0 + \left(K \alpha^2+ \frac{\Delta \alpha}{dT_c/dP} \right) T, & T \leq T_c,
     \end{cases}
\end{align}
where $C_0=T\left(\partial S_0 / \partial T\right)_P$ is the volumetric heat capacity of the background. 

\subsection{Thermodynamic properties of the background} \label{sec:background}

Equations~(\ref{eq:volume}), (\ref{eq:entropy2}), and (\ref{eq:specific heat capacity}) depend on the thermodynamic properties of the background $V_0, S_0$ and $C_0$, which are formally derived from the background free-energy density $G_0$. While in Landau theory such contributions are unimportant for describing thermodynamic properties associated solely with the phase transition, they are important in our model for calculating BC effects.   

We determine $G_0$ by first assuming that $V_0$ is independent of temperature and pressure; second, by parametrizing $C_0$ with a function linear in temperature and independent of pressure; and third, by integrating $V_0=\left(\partial G_0 / \partial P\right)_{T}$ and $C_0=-T \left(\partial^2 G_0 / \partial T^2 \right)_P$. Explicit expressions for $C_0$, $G_0$, and $S_0$ are given in Sec.~\ref{sec:C60}.

Our parametrization of $C_0$ is justified by noting that a linear function in temperature approximates well the ambient-pressure heat capacity  of the disordered phase of BC PCs~\cite{Lloveras2019a, Junning2020a, Diky2001a, Charapennikau2002a, Charapennikau2003a}; and that the assumption of $V_0$ independent of temperature implies that $C_0$ is independent of pressure, as follows from the thermodynamic relation $\left(\partial C_0 / \partial P\right)_T = -T \left(\partial^2 V_0 / \partial T^2\right)_P$. We note that $V_0$ independent of temperature also implies that $S_0$ does not depend on pressure, as follows from the Maxwell relation $\left(\partial S_0 / \partial P\right)_T =  - \left(\partial V_0 / \partial T\right)_P$. 
 
We justify the assumption of $V_0$ independent of temperature and pressure by the overall agreement between our model and experiments, most notably, between the predicted and observed thermal expansion, see Sec.~\ref{sec:comparison} and Fig.~\ref{fig:Comparison}\,(c). Such assumption holds in the vicinity of the phase transition, and would fail away from it as the thermal expansion of the individual plastic and crystal phases becomes non-linear in temperature~\cite{Li1994a}.
\vspace{-0.39cm}
\subsection{BC effects} \label{sec:BCEs}
\vspace{-0.39cm}
We calculate isothermal changes in entropy $ \Delta S(T,P)$ and adiabatic changes in temperature $\Delta T(T_s,P)$ 
on compression~($0 \to P$) and decompression ($P \to 0$)
following the standard procedure~\cite{Lloveras2023a},
\begin{align}
    \label{eq:DS}
    \Delta S(T,P)= \begin{cases} S(T,P)-S(T,0), & 0 \to P, \\ S(T,0)-S(T,P), & P \to 0,  \end{cases}
\end{align}
and,
\begin{align}
    \label{eq:DT}
    \Delta T(T_s,P)= \begin{cases} T(S,P)-T_s(S,0), & 0 \to P, \\ T(S,0)-T_s(S,P), & P \to 0,  \end{cases} 
\end{align}
where $T_s$ is the starting temperature. In general, 
both $\Delta S(T,P)$ and $\Delta T(T_s,P)$ implicitly involve the background entropy density $S_0$, see Eq.~(\ref{eq:entropy2}). However, given that $S_0$ is independent of pressure within our approximations (Sec.~\ref{sec:background}), only $\Delta T(T_s,P)$ depends on $S_0$.

\begin{figure}[H]
    \centering
    \includegraphics[width=0.75\columnwidth]{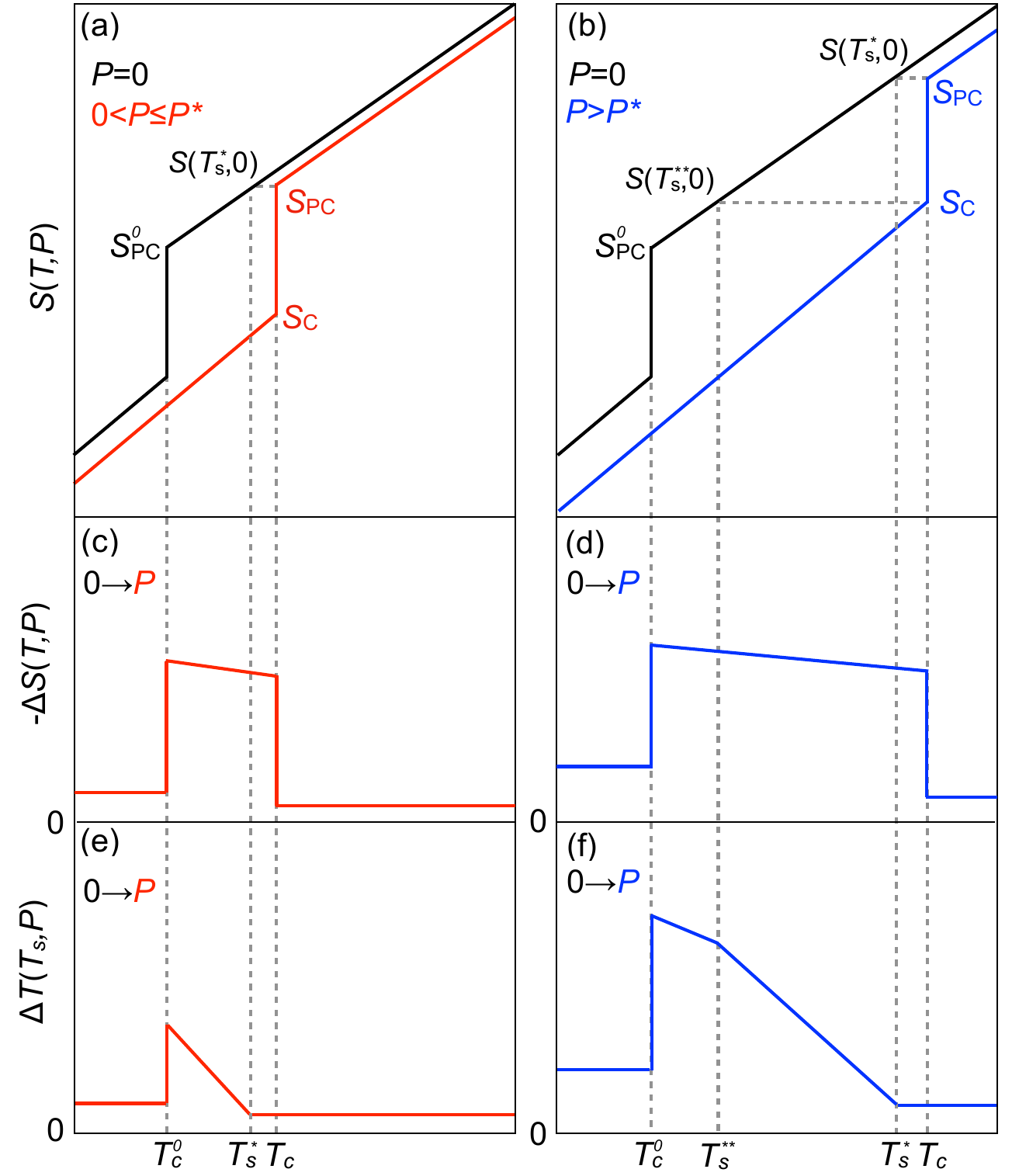}
    \caption{Schematics of predicted BC effects on compression for a material with $dT_c/dP>0$. Entropy density $S(T,P)$ for pressures (a) below and (b) above the threshold pressure $P^*$ at which $S^0_{\rm PC}=S_{\rm C}$; isothermal changes in entropy $\Delta S(T,P)$ for pressures (c) below and (d) above $P^*$; and adiabatic changes in temperature $\Delta T(T_s,P)$ for pressures (e) below and (f) above $P^*$.  For $P\leq P^*$, $T_c^0, T_c$ and $T_s^*$ determine the temperature regions associated with the 
    BC effects from the individual plastic and crystal phases and those involving the phase transition. For $P >P^*$, $T_s^{**}$ determines an additional temperature region associated with BC effects involving the phase transition. 
    Here, $S_{\rm PC}^0$ is the value of the entropy density at ambient pressure as $T$ approaches $T_c^0$ from above; $S_{\rm PC}$ is the value of the entropy density at finite pressure as $T$ approaches $T_c$ from above; $S_{\rm C}$ is the value of the entropy density at finite pressure as $T$ approaches $T_c$ from below;
    $T_s^{*}$ is the starting temperature at which $S(T_s^{*},0)=S_{\rm PC}$; and $T_s^{**}$ is the starting temperature at which $S(T_s^{**},0)=S_{\rm C}$.}
    \label{fig:BC_Schematics}
\end{figure}

\section{Results and discussion} \label{sec:RandD}

Our model predicts that the onset of long-range orientational ordering leads to (i) a change $\Delta \alpha$ in the coefficient of thermal expansion (Eqs.~(\ref{eq:volume}) and (\ref{eq:Dalpha})), (ii) a change in volume $\Delta V_t$ (Eqs.~(\ref{eq:volume}) and~(\ref{eq:DVt})),  (iii) a change in entropy $\Delta S_t$ (Eqs.~(\ref{eq:entropy2}) and~(\ref{eq:DSt})), and (iv) a linear temperature-pressure phase boundary (Eq.~(\ref{eq:Tc})), which is in agreement with experiments~\cite{Lloveras2019a, Li2019a, Meijer2023a, Aznar2020a, Salvatori2023b, Aznar2021a, Salvatori2022a, Zeng2024a, Junning2020a, Zhang2022a, Ren2022a, Zhang2023a, Zhang2023b}.  We note that $\Delta \alpha $, $\Delta V_t$, and the rate $dT_c/dP$ (Eq.~(\ref{eq:dTcdP}))  depend on the strain-orientation coupling. The main discrepancy between our model and experiments ~\cite{Lloveras2019a, Aznar2020a, Salvatori2023b, Aznar2021a, Salvatori2022a, Junning2020a}), is that the predicted  $\Delta V_t$ and $\Delta S_t$ are independent of pressure. This shortcoming stems from neglecting non-linear invariants beyond those we have considered in the strain-orientation coupling energy $G_{\epsilon Q}$, which yield the change $\Delta A_t$ independent of $P$~(Eq.~(\ref{eq:OPjump})).

We now turn to the BC effects predicted by our model, which are schematically shown in Fig.~\ref{fig:BC_Schematics}  on compression for a material with $dT_c/dP>0$. We first note that the BC response depends on a threshold pressure $P^*$ at which $S_{\rm PC}^0 = S_{\rm C}$, where $S_{\rm PC}^0 \equiv S_0\left(T_c^0\right)$ is the value of the entropy density at ambient pressure as $T$ approaches $T_c^0$ from above, and $S_{\rm C} \equiv S_0\left(T_c\right)-\left(1-K\alpha dT_c/dP\right)\alpha P+\Delta S_t$ is the value of the entropy density at finite pressure as $T$ approaches $T_c$ from below, see Figs.~\ref{fig:BC_Schematics}\,(a), \ref{fig:BC_Schematics}\,(b), and Eq.~(\ref{eq:entropy2}). For $P \leq P^*$, the temperature-independent isothermal changes in $\Delta S(T,P)$ result solely from the thermal expansion of the individual plastic ($T > T_c$) and crystal phases ($T < T_c^0$), while the temperature-dependent part in $\Delta S(T,P)$ originates from the phase transition ($T_c^0 < T < T_c$), as shown in Fig.~\ref{fig:BC_Schematics}\,(c). For $P>P^*$, $\Delta S(T,P)$ increases in magnitude while retaining its overall shape, see Fig.~\ref{fig:BC_Schematics}\,(d). 

For the adiabatic changes $\Delta T(T_s,P)$ at $P \leq P^*$, the temperature-independent responses associated with the individual plastic and crystal phases occur, respectively, for $T_s < T_c^0$ and $T_s > T_{s}^{*}$, while the temperature-dependent changes involving the phase transition occur for $T_c^0 < T_s < T_{s}^{*}$ (Fig.~\ref{fig:BC_Schematics}\,(e)). Here, $T_s^{*}$ is the starting temperature at which  $S(T_s^{*},0)=S_{\rm PC}$, where  $S_{\rm PC} \equiv S_0(T_c)-\left(1-K\alpha dT_c/dP\right)\alpha P$ is the value of the entropy density at finite pressure as $T$ approaches $T_c$ from above, see Fig.~\ref{fig:BC_Schematics}\,(e) and Eq.~(\ref{eq:entropy2}). For $P>P^*$, $\Delta T(T_s,P)$ exhibits two temperature-dependent regions at $T_c^0 < T_s < T_s^{**}$ and $T_c^* < T_s < T_s^{*}$, where $T_s^{**}$ is the starting temperature at which $S(T_s^{**},0)=S_{\rm C}$, see Fig.~\ref{fig:BC_Schematics}\,(f). We note that a similar analysis for a material with $dT_c/dP<0$ would yield inverse BC effects~\cite{Lloveras2023a}. 

\subsection{Case example: solid \texorpdfstring{C$_{60}$}{TEXT}} 

\subsubsection{Fits to experiments} \label{sec:C60}

 At ambient pressure and above  $\sim 260\,$K, the C$_{60}$ molecules in fullerite are orientationally disordered and their centre of masses lie in a face-centered cubic lattice~\cite{Moret2005a}. Upon cooling, fullerite undergoes a first-order phase transition to a simple-cubic lattice structure in which the molecular orientations are partially ordered~\cite{Tycko1991a, Heiney1991a},with a relative volume reduction of about $1\,\%$~\cite{David1992a}. For applied hydrostatic pressures in the range in which BC effects have been experimentally observed ($0-0.60\,{\rm GPa}$)~\cite{Junning2020a}, the transition temperature increases up to $\sim  360\,{\rm K}$. The orientational OP of the phase transition is a tetrahexacontapole (64-pole)~\cite{Harris1992a}, and while moments of lower order such as the quadrupole ${\bm Q}$ vanish due to molecular symmetry, we still use our model, as Eq.~(\ref{eq:renormalizedG}) is isomorphic to the free-energy density  of fullerite~\cite{Harris1992a}.

 We obtain model parameters by performing the following fits to experimental data~\cite{Junning2020a}: 
$T_c^0$ and $dT_c/dP$ are obtained by fitting 
Eq.~(\ref{eq:Tc}) to the temperature-pressure boundary on heating; $\alpha$, $\Delta \alpha$  and $\Delta V_t$ are obtained by fitting Eq.~(\ref{eq:volume}) with $V_0=1$ to the observed volume at ambient pressure.  The results are as follows: $T_c^0 =263\,{\rm K}, dT_c/dP=165\,{\rm K\,GPa^{-1}}, \alpha =42\times 10^{-6}\,{\rm K^{-1}}, \Delta \alpha = 36\times 10^{-6}\,{\rm K^{-1}}$, and $\Delta V_t=-0.012$. We take $K=18.1\,{\rm GPa}$ from the measured bulk modulus at ambient conditions~\cite{Duclos1991a}. 

By using the procedure described in Sec.~\ref{sec:BCEs}, and fitting Eq.~(\ref{eq:specific heat capacity}) to the observed specific heat capacity of the disordered phase~\cite{Diky2001a}, we obtain the background specific heat capacity $C_0= -163.13\,{\rm J \, K^{-1} \,kg^{-1}} + 2.94\,{\rm J \, K^{-2} \, kg^{-1}} T$, which yields the background free-energy  $G_0=-163.13\,{\rm J \, K^{-1} \,kg^{-1}} \left[1 + \ln\left(T/240\,{\rm K}\right) \right] T -  1.47\,{\rm J \, K^{-2} \, kg^{-1}} (T-240\,{\rm K})^2 + \rho^{-1} P$, and the background entropy $S_0 =  -163.13\,{\rm J \, K^{-1} \,kg^{-1}} +  2.94\,{\rm J \, K^{-2} \, kg^{-1}} (T-240\,{\rm K})$, where $\rho = 1.7\times 10^3\,{\rm kg \, m^{-3}}$ is the density, and $240\,{\rm K}$ is the temperature at which we fit the value of the ambient-pressure entropy as reported in Ref.~\cite{Junning2020a}, i.e. $S(240\,{\rm K},0)=193\,{\rm J\,K^{-1}\,kg^{-1}}$.
\vspace{-0.5cm}
\subsubsection{Comparison to experiments} 
\vspace{-0.4cm}
\label{sec:comparison}

\begin{figure}[H]
    \centering
    \includegraphics[width=\columnwidth]{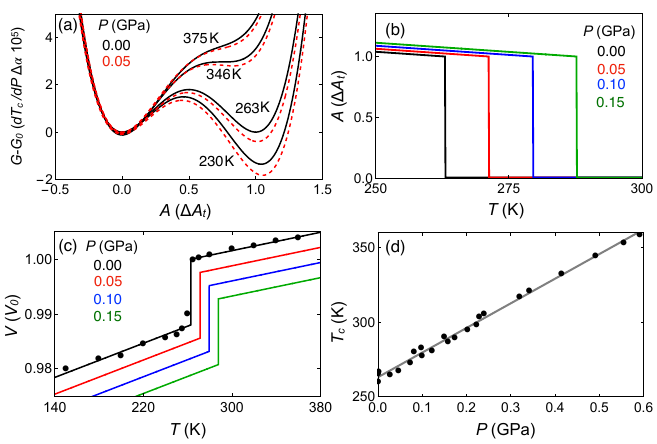}
    \caption{Case example for solid C$_{60}$. Calculated (a) excess free-energy density, (b) scalar order parameter, (c) relative volume, and (d) temperature-pressure phase boundary.  Experimental data ($\bullet$) taken from Ref.~\cite{Junning2020a}.}
    \label{fig:Comparison} 
\end{figure}

We now turn to the results predicted by our model using the above parametrization. Figures~\ref{fig:Comparison}\,(a) and (b) show, respectively, the excess free-energy density $G-G_0$~(Eq.~(\ref{eq:renormalizedG})) and the scalar OP $A(T,P)$~(Eq.~(\ref{eq:Aexpansion})) at several pressures, both of which are typical for orientationally disordered systems~\cite{Selinger2015a}.
Figures~\ref{fig:Comparison}\,(c) and ~\ref{fig:Comparison}\,(d) show, respectively, the predicted relative volume at several pressures (Eq.~(\ref{eq:volume})), and the temperature-pressure phase boundary (Eq.~(\ref{eq:Tc})), both of which are in excellent agreement with experiments~\cite{Junning2020a}.

\begin{figure}[H]
    \centering
    \includegraphics[width=0.5\columnwidth]{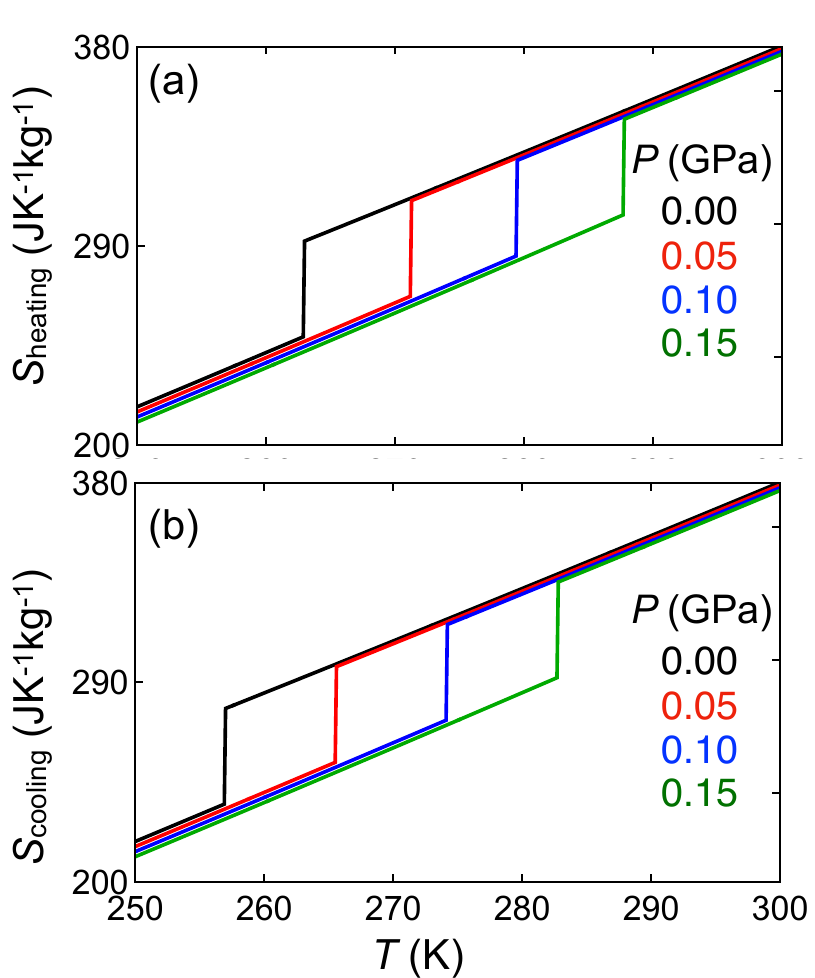}
    \caption{Calculated temperature dependence of the entropy on (a) heating and (b) cooling at several pressures for solid C$_{60}$.}
    \label{fig:entropies}
\end{figure}

BC effects in solid C$_{60}$ were determined on compression  and decompression from the entropies on cooling and heating ramps, respectively~\cite{Junning2020a}. Therefore, for the purposes of comparison, we calculate the isothermal entropy changes accordingly~\cite{Lloveras2023a}, i.e. $\Delta S(T,P)= S_{\rm cooling}(T,P)-S_{\rm cooling}(T,0)$ for $0 \to P$, and $\Delta S(T,P)= S_{\rm heating}(T,0)-S_{\rm heating}(T,P)$ for $P \to 0$; as well as the adiabatic changes in temperature, i.e. $\Delta T(T_s,P)= T(S_{\rm cooling},P)-T_s(S_{\rm cooling},0)$ for $0 \to P$, and $\Delta T(T_s,P)= T(S_{\rm heating},0)-T_s(S_{\rm heating},P)$ for $P \to 0$. Here, $S_{\rm heating}(T,P)$ is the predicted entropy  on heating (Fig.~\ref{fig:entropies}\,(a)) calculated from Eq.~(\ref{eq:entropy2}) with the above parametrization, and $S_{\rm cooling}(T,P)$ is the predicted entropy on cooling (Fig.~\ref{fig:entropies}\,(b)) calculated from Eq.~(\ref{eq:entropy2}) with the above parametrization, except that $T_c^0 =257\,{\rm K}$ and $dT_c/dP=172\,{\rm K\,GPa^{-1}}$, which we obtained by fitting Eq.~(\ref{eq:Tc}) to the experimental temperature-pressure phase boundary on cooling~\cite{Junning2020a}. 

\begin{figure}[H]
    \centering
    \includegraphics[width=\columnwidth]{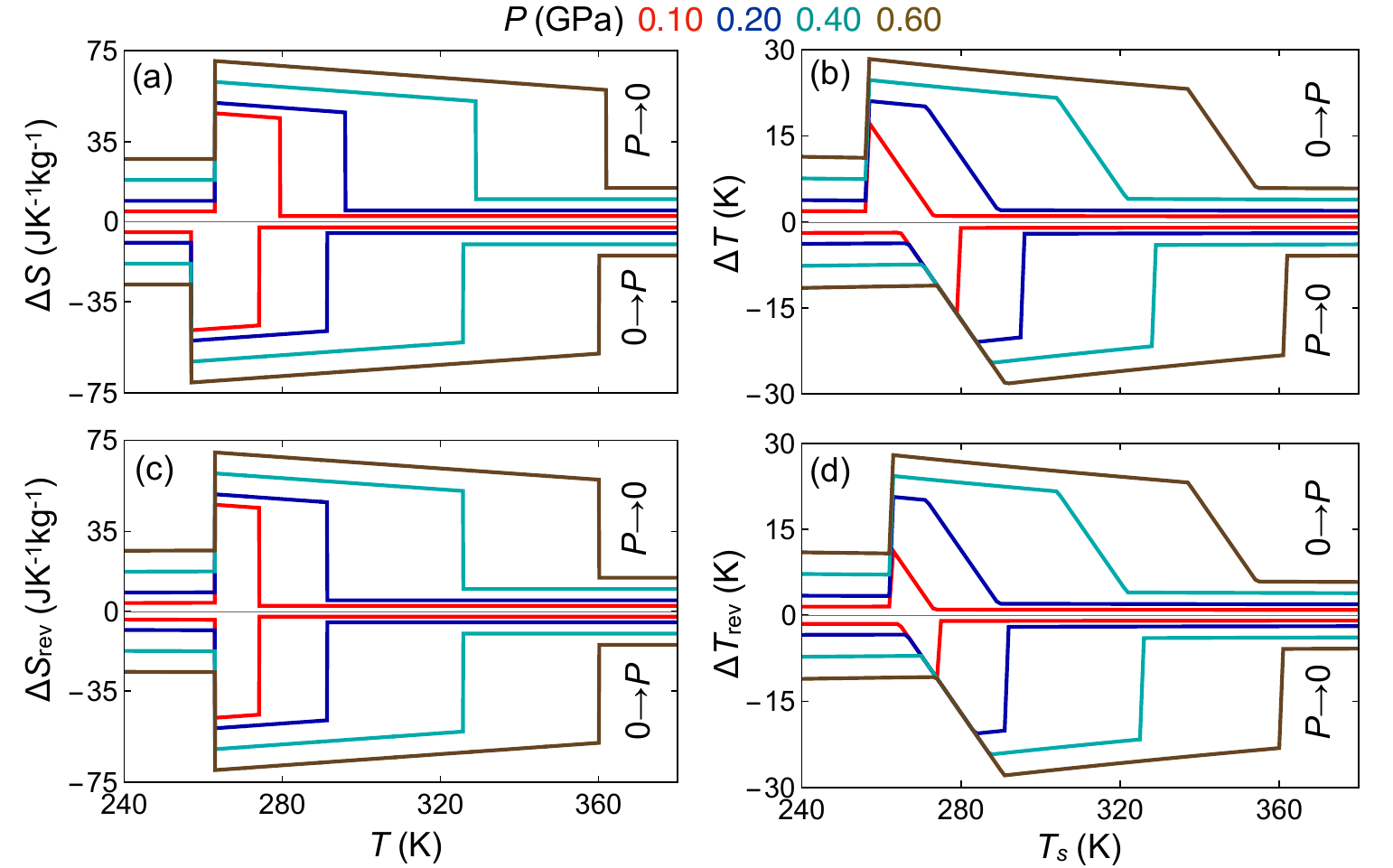}
    \caption{Calculated (a) isothermal changes in entropy on compression and decompression, and (b) associated reversible entropy changes. Calculated (c) adiabatic changes in temperature on compression and decompression, and (d) associated reversible changes in temperature for solid C$_{60}$.}
    \label{fig:bce}
\end{figure}

\begin{figure}[H]
    \centering
    \includegraphics[width=0.5\columnwidth]{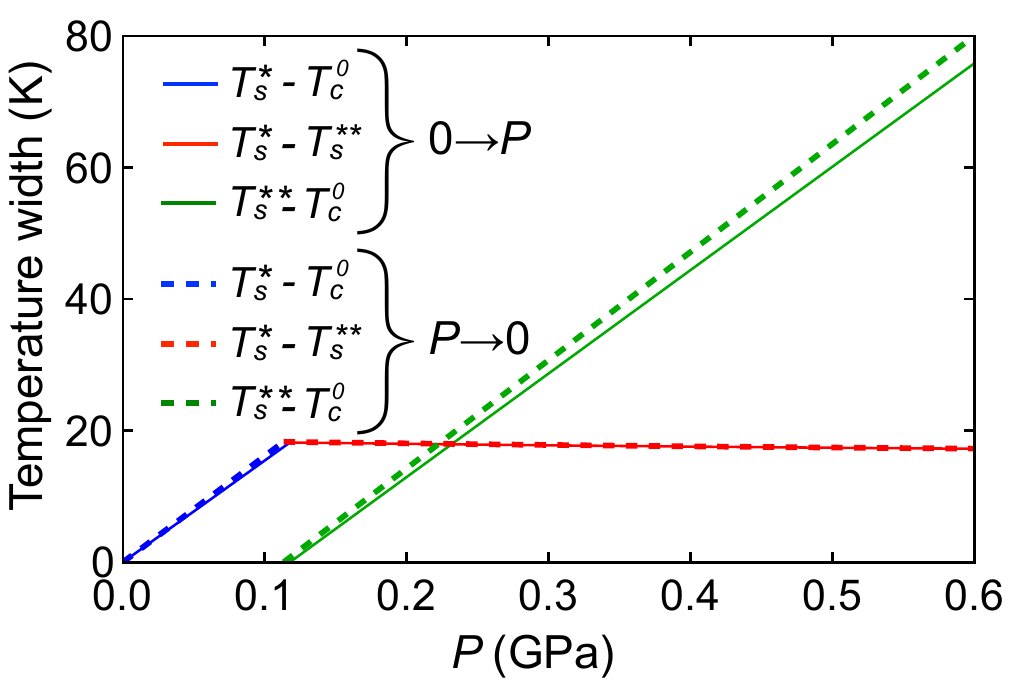}
    \caption{Calculated temperature spans on compression and decompression of the characteristic regions in the adiabatic temperature changes $\Delta T(T_s,P)$ for solid C$_{60}$.}
    \label{fig:DTregions}
\end{figure}

We now assess the reversibility of the BC response~\cite{Aznar2020a}. In order to account for the transition hysteresis, we calculate reversible isothermal changes in entropy $\Delta S_{\rm rev}$ and reversible adiabatic changes in temperature $\Delta T_{\rm rev}$ using the standard procedure~\cite{Lloveras2023a}, namely, $\Delta S_{\rm rev}(T,P) = S_{\rm heating}(T,0)-S_{\rm cooling}(T,P)$ for $0 \to P$, $\Delta S_{\rm rev}(T,P) = S_{\rm heating}(T,0)-S_{\rm cooling}(T,P)$ for $P \to 0$, $\Delta T_{\rm rev}(T_s,P)= T(S_{\rm cooling},P)-T_s(S_{\rm heating},0)$ for $0 \to P$, and $\Delta T_{\rm rev}(T_s,P)=T(S_{\rm heating},0)-T_s(S_{\rm cooling},P)$ for $P \to 0$. The results are shown in Fig.~\ref{fig:bce}. Our model captures the characteristic shape of the BC effects
in the individual plastic and crystal phases, as well as the shape associated with the phase transition. On compression and decompression, we obtain identical threshold pressures ($P^*=0.12\,{\rm GPa}$), and nearly identical temperature spans $T_s^*-T_c^0, T_s^*-T_s^{**}$ and $T_s^{**}-T_c^0$ (Fig.~\ref{fig:DTregions}), which is a consequence of the small hysteresis ($\lesssim 6\,$K).

We find that our model overestimates the magnitude of the BC effects reported in Ref.~\cite{Junning2020a}. We attribute such discrepancy to the predicted transition entropy $\norm{\Delta S_t}= 43\,{\rm J  K^{-1} kg^{-1}}$, which is pressure independent and greater than the value reported in Ref.~\cite{Junning2020a} ($\sim 27\,{\rm J  K^{-1} kg^{-1}}$) at ambient pressure. However, as noted in Ref.~\cite{Junning2020a}, the observed transition entropy is characteristic of powdered samples~\cite{Heiney1991a, Samara1993a}, which are smaller than in single crystals ($\sim 63\,{\rm J  K^{-1} kg^{-1}}$~\cite{Miyazaki99a}). This suggests that significantly larger BC effects could be reached in solid C$_{60}$. 
Indeed, using $\Delta S_t=-63\,{\rm J  K^{-1} kg^{-1}}$, we predict peak values $|\Delta S | \simeq  90\,{\rm J K^{-1} kg^{-1}}$ and $|\Delta T| \simeq 35\,$K for $0.6$\,GPa at $265\,$K.

\section{Summary and Conclusions}

We have proposed a Landau theory for PCs in which orientational ordering is the sole driver of the plastic-to-crystal phase transition, thus enabling predictions of associated conventional and inverse BC effects. The key components of our theory are a multiple-moment OP that characterizes the collective alignment of the constituent molecules, its coupling to elastic strains, and the thermal expansion. We have illustrated these ideas with the simplest non-trivial model in which the orientational ordering is described by a quadrupolar OP. Our theory predicts the general behavior found in several classes of BC PCs~\cite{Lloveras2019a, Salvatori2022a, Li2020a} and related orientationally disordered crystals~\cite{Ren2022a, Zhang2023a, Zhang2023b}, while also providing a physical framework for understanding their thermodynamic properties. 

The most significant limitation of our model is the absence of pressure dependency in the transition entropy change $\Delta S_t$ and the transition volume change $\Delta V_t$, which we attributed to neglecting non-linear invariants beyond those we considered in the orientation-strain coupling energy. Such interactions will add to a more precise description of the plastic-to-crystal transition and associated BC effects, but will not change the main predictions of our theory. 

We have used solid C$_{60}$ as a case example and obtained excellent agreement with the observed thermal expansion and the temperature-pressure phase boundary, and good agreement with the observed BC effects. We have identified the temperature regions in the BC response associated with  the individual plastic and crystal phases, as well as those involving the phase transition. For C$_{60}$ single crystals, we anticipate peak isothermal entropy changes and peak adiabatic temperature changes as large as $\sim 90 \,{\rm J K^{-1} kg^{-1}}$ and $\sim 35 \,{\rm K}$, respectively, using $0.60\,$GPa at $265\,$K. 

Finally, we note that our work provides a starting point for developing theories for electrolyte~\cite{Zeng2024a} and ferroelectric~\cite{Harada2016a} PCs where the corresponding ionic diffusion and polarization likely contribute to the BC effects.

\newpage

\section{Acknowledgements} G. G. G.-V acknowledges helpful feedback and discussions with Pol Lloveras, Enric Stern-Taulats, and Melony Dilshad. G. G. G.-V and R. M.-D. acknowledge support from the Vice-rectory for Research at the University of Costa Rica. G. G. G.-V is grateful to Churchill College at the University of Cambridge for hospitality. X. M. acknowledges funding from the UK EPSRC (Grant Number EP/V042262/1‌ and EP/M003752/1), ERC Starting Grant (no. 680032) and the Royal Society.

\newpage


%

\end{document}